\documentstyle[aps,prl,preprint]{revtex}

\def\r{\hat\rho}
\def\R#1{\hat\rho^{(#1)}}
\def\H#1{{\cal H}^{\otimes #1}}
\def\tr{{\rm tr}}

\def\sx{\hat\sigma_x}
\def\sy{\hat\sigma_y}
\def\sz{\hat\sigma_z}
\def\rxyz{\hat\rho_{x,y,z}}
\def\intdr{\int\!\! d\hat\rho\,}

\begin{document}

\tightenlines

\title{Quantum Bayes rule}

\author{R\"udiger Schack$^1$, Todd A. Brun$^2$, and Carlton M. Caves$^3$ \\
$^1${\it Department of Mathematics, Royal Holloway, University of London,\\
Egham, Surrey TW20 0EX, UK} \\
$^2${\it Physics Department, Carnegie Mellon University,
Pittsburgh, PA 15213, USA} \\
$^3${\it Center for Advanced Studies, Department of Physics and Astronomy,} \\
{\it University of New Mexico, Albuquerque, NM 87131, USA} }

\date{\today}

\maketitle

\begin{abstract}
  We state a quantum version of Bayes's rule for statistical inference and
  give a simple general derivation within the framework of generalized
  measurements.  The rule can be applied to measurements on $N$ copies of a
  system if the initial state of the $N$ copies is \textit{exchangeable}.  As
  an illustration, we apply the rule to $N$ qubits.  Finally, we show that
  quantum state estimates derived via the principle of maximum entropy are
  fundamentally different from those obtained via the quantum Bayes rule.
\end{abstract}


During the last decade, interest in Bayesian methods of statistical inference
has increased considerably \cite{Bernardo1994,Malakoff1999}. At the heart of 
the Bayesian approach is Bayes's rule, which indicates how to update a state 
of knowledge in the light of new data. The simplest form of the rule is
\begin{equation}
p(H|D) = {p(D|H) p(H)\over p(D)} \;,
\label{eq:classicalbayes}
\end{equation}
where $p(D|H)$ is the probability for the data $D$ given a hypothesis $H$, 
$p(H)$ is the \textit{prior} probability that the hypothesis is true, $p(H|D)$ 
is the \textit{posterior} probability that the hypothesis is true given the 
data, and $p(D)=\sum_H p(D|H)p(H)$ is the probability for the data averaged
over all hypotheses. The conceptual simplicity of Bayes's rule is a major 
strength of the Bayesian approach.

The problems of statistical inference and state estimation are of central
importance in quantum information theory.  After the early pioneering work on
quantum inference \cite{Sykora1974,Helstrom1976,Holevo1982,Jones1991a} and
quantum state tomography \cite{Vogel1989b,Smithey1993,Leonhardt1995}, a large 
amount of work has been done on the subject (see, e.g., Refs.\ 
\cite{Massar1995,Hradil1997,Slater1997,Derka1998,Buzek1998b,%
Tarrach1999,Banaszek2000,Gill2000a,Lemm2000a,Paini-0002}).  In many of
the cited papers, a quantum version of Bayes's rule is used either implicitly
or explicitly. Jones \cite{Jones1991a} has derived a quantum Bayes rule for
pure states only. In this paper, we derive a general rule, valid both for pure
and mixed states, and give a precise condition for its validity.

We consider the following general inference problem.
Let ${\cal H}$ be the Hilbert space of a quantum system. The Hilbert space of
$N$ copies of the system is given by the $N$-fold tensor product, $\H N$.
Suppose one is given a (prior) state $\R{M+N}$ on $\H{(M+N)}$ and the results
of measurements on $M$ subsystems. The task is to find the (posterior)
state of the remaining $N$ subsystems conditioned on the measurement results.
The problem is in principle completely solved by the theory of generalized
measurements \cite{Kraus1983}, which prescribes the state of the total system
after the measurement. There is no room in quantum theory for an
additional independent inference principle; any inference rule must be
derivable from the basic theory.

An arbitrary measurement
on $M$ subsystems is described by a set of completely positive,
trace-decreasing operations, $\{{\cal F}_k\}$, which act on the
selected $M$ subsystems. The measurement result is $k$ with
probability 
\begin{equation}
p_k=\tr\big[{\cal F}_k\big(\R{M+N}\big)\big] \;.
\label{eq:pkdef}
\end{equation}
Since the operations ${\cal F}_k$ are completely positive, they can be expressed in the form 
\begin{equation}
{\cal F}_k\big(\R{M+N}\big) = \sum_l (\hat A_{kl}\otimes\hat1)\,\R{M+N}\,
        (\hat A^\dagger_{kl}\otimes\hat1) \;,
\label{eq:Fk}
\end{equation}
where the $\hat A_{kl}$ are arbitrary operators acting on the selected
$M$ subsystems. The probabilities $p_k$ can thus be rewritten as
\begin{equation}
p_k=\tr\big[(\hat E_k\otimes\hat1)\R{M+N}\big] 
=\tr_M\big(\hat E_k\R{M}\big) \;,
\label{eq:pk}
\end{equation}
where 
\begin{equation}
\hat E_k = \sum_l \hat A^\dagger_{kl} \hat A_{kl} 
\end{equation}
is a positive semidefinite operator and 
\begin{equation}
\sum_k \hat E_k = \hat 1 \;;
\end{equation}
i.e., the set $\{\hat E_k\}$ forms a positive operator valued measure (POVM).
In the last form of Eq.~(\ref{eq:pk}), $\R{M}$ is the prior marginal density 
operator of the measured subsystems, and $\tr_M$ denotes a trace over the 
measured subsystems.

If the measurement result is $k$, the (normalized) state of all $M+N$ systems 
after the measurement is
\begin{equation}
\R{M+N}_k = {1\over p_k} {\cal F}_k\big(\R{M+N}\big) \;,
\end{equation}
where ${\cal F}_k(\R{M+N})$, given in Eq.~(\ref{eq:Fk}),
is the unnormalized state conditioned on 
measurement outcome $k$.  Performing a partial trace over the selected 
$M$ subsystems yields the posterior state of the remaining $N$ subsystems,
\begin{eqnarray}
\R N_k = \tr_M\big(\,\R{M+N}_k\big) \;.
\label{partial_trace}
\end{eqnarray}
An exact quantum analogue of the classical Bayes rule would write this 
posterior state as a mixture in which the updating as a consequence of
obtaining result~$k$ (the ``data'') would appear in the probabilities in 
the mixture, but not in the density operators that contribute to the 
mixture.  Classically it is possible to obtain information about a 
system without disturbing it, while quantum mechanically it is not; hence, 
Eq.~(\ref{partial_trace}) must include both updating due to the information 
acquired and due to the disturbing effects of the measurement.  In general, 
this only takes the form of the classical Bayes rule if the measured and 
unmeasured systems in Eqs.~(\ref{eq:pkdef})--(\ref{partial_trace}) are 
initially unentangled.

Notice also that for a product prior, 
\begin{equation}
\R{M+N}=\r_0^{\otimes(M+N)}\equiv
\underbrace{\r_0\otimes\cdots\otimes\r_0}_{\mbox{$M+N$ terms}}\;,
\label{eq:productprior}
\end{equation}
where $\hat\rho_0$ is some state on ${\cal H}$, the posterior state is 
$\R N_k=\r_0^{\otimes N}$, irrespective of the measurement result.  No 
learning from data is possible for product priors.  This shows in particular 
that the totally mixed state for $M+N$ subsystems, which is both a product 
state and the state of maximum entropy on $\H{(M+N)}$, does not allow 
learning from measured data.

In many practical situations, one can restrict attention to prior
states of the form
\begin{equation}
\R N = \intdr p(\r) \r^{\otimes N} \;,
\label{eq:exch}
\end{equation}
where $d\r$ is a measure on density operator space and $p(\r)$ is a
normalized generating function, $\intdr p(\r)=1$. Prior states of the form
(\ref{eq:exch}) arise, e.g., if each subsystem is prepared in the same,
unknown way, as in quantum state tomography. A state of 
$N$ subsystems, $\R N$, can be expressed in the form (\ref{eq:exch}) if 
and only if it is {\it exchangeable}, i.e., if (i) it is invariant under 
permutations of the subsystems and (ii) for any $M>0$, there is a state 
$\R{N+M}$ of $N+M$ subsystems that is invariant under permutations of 
the subsystems and that satisfies $\R N=\tr_M\big(\R{N+M}\big)$ 
\cite{Stormer1969,Hudson1976}. 
The expansion (\ref{eq:exch}) is then unique. This 
is the quantum version of the fundamental representation theorem due to 
de Finetti \cite{DeFinetti1990}; for an elementary proof of the quantum 
theorem see Ref.~\cite{Caves2000a}.

The significance of part~(ii) of the definition of exchangeability
given above is illustrated by the GHZ state 
$\hat\rho_{\rm GHZ}=|\psi_{\rm GHZ}\rangle\langle\psi_{\rm GHZ}|$, 
where $|\psi_{\rm GHZ}\rangle=(|000\rangle+|111\rangle)/\sqrt2$. This
three-particle state is invariant under permutations of the three
subsystems, but it is clear that $\hat\rho_{\rm GHZ}$ cannot be obtained
by a partial trace from a four-particle state that is invariant under
permutations of all four particles. The GHZ state is thus not
exchangeable, in accordance with the fact that it cannot be
written in the form~(\ref{eq:exch}).

If the condition of exchangeability is fulfilled, the question of
finding a suitable prior state reduces to finding a suitable prior
measure $p(\hat\rho)d\rho$ in the expansion (\ref{eq:exch}). Much work 
has been done on suitable prior measures on density operator space (see, 
e.g., \cite{Slater1997,Bures1969,Braunstein1994a,Zyczkowski1998b}). As in
the classical theory of inference \cite{Bernardo1994}, there exists no
unique choice of prior measure; different kinds of prior information
lead to different prior measures.

The rule of inference, however, becomes extremely simple if the prior
state is of the form (\ref{eq:exch}). In this case, we show below
that if a measurement performed on the first subsystem yields result~$k$, 
the posterior state of the remaining $N-1$ subsystems is given by
\begin{equation}
\R{N-1} = \intdr  p(\r|k) \r^{\otimes(N-1)} \;,
\label{eq:N-1}
\end{equation}
where
\begin{equation}
p(\r|k) = { p(k|\r)p(\r) \over p_k} \;.
\label{eq:bayes}
\end{equation}
Here $p(k|\r) = \tr(\hat E_k\r)$ is the probability of
obtaining the measurement result $k$ for a single subsystem, given
that the state of the single subsystem is $\r$, and $p_k=\intdr p(k|\r)p(\r)$
is the average probability of obtaining $k$. This is the quantum
Bayes rule; it is completely analogous to the classical 
rule~(\ref{eq:classicalbayes}). 

In the special case that the integration in Eq.~(\ref{eq:exch}) is restricted
to pure states, the rule (\ref{eq:bayes}) has been derived by Jones
\cite{Jones1991a} and applied to purifications of mixed states by Bu\v{z}ek
{\it et al.} \cite{Buzek1998b}. Tarrach and Vidal \cite{Tarrach1999} have
used Eq.~(\ref{eq:bayes}) to find optimal measurements on $N$ copies of a
system, identically prepared in an unknown mixed state by some preparation
device. To our knowledge, Eq.~(\ref{eq:bayes}) has not been derived in the
general context considered here.

If measurements are performed on several subsystems individually, the rule
(\ref{eq:bayes}) can be simply iterated. Although the situation
considered here, where measurements are done one subsystem at a
time, is in practice the most important, it is straightforward to
generalize the rule to the case of collective measurements on several
subsystems. 

Strictly speaking, the generating function $p(\r)$ should not be called a
probability---after all, a mixed state $\r$ is itself a summary of incomplete
knowledge about a subsystem. Nevertheless, the content of the quantum Bayes
rule (\ref{eq:bayes}) is that the functions $p(\r)$ and $p(\r|k)$ can be used
as if they were a prior probability and a conditional posterior probability
for density operators. This interpretation is obviously appropriate in the case 
that the exchangeable state (\ref{eq:exch}) is known to have arisen from an
experiment in which each subsystem is prepared in the same unknown state,
with $p(\r)$ then being the probability that this unknown state is~$\r$. 

To derive the rule (\ref{eq:bayes}), we denote by  $\{{\cal F}_k\}$ the 
set of completely positive, trace-decreasing operations which describe 
the measurement on the first subsystem. The result of the measurement
is $k$ with probability
\begin{equation}
p_k=\tr\big[{\cal F}_k\big(\R N\big)\big] = \intdr  p(k|\r) p(\r) \;.
\end{equation}
If the measurement result is $k$, the state of all $N$ subsystems after the
measurement is 
\begin{equation}
\R N_k
 = {1\over p_k} \intdr p(\r) {\cal F}_k(\r)\otimes\r^{\otimes(N-1)} \;,
\end{equation}
where, by a slight abuse of notation, we denote by ${\cal F}_k(\r)$
the unrenormalized state of a single subsystem with premeasurement
state $\hat\rho$ conditioned on the measurement result $k$.  A partial
trace over the first subsystem gives the state of the remaining $N-1$
subsystems,
\begin{eqnarray}
\R {N-1}_k &=& \tr_1\big(\R N_k\big) \nonumber\\
&=&{1\over p_k}\intdr p(\r)\tr[{\cal F}_k(\r)]\r^{\otimes(N-1)} \nonumber\\
&=&{1\over p_k}\intdr p(\r)\tr(\hat E_k\r)\r^{\otimes(N-1)} \nonumber\\
&=&\intdr {p(\r)p(k|\r)\over p_k}\r^{\otimes(N-1)} \nonumber\\
&=&\intdr p(\r|k)\r^{\otimes(N-1)} \;,
\end{eqnarray}
where in the last line we have substituted $p(\r|k)$ for the right-hand side
of Eq.~(\ref{eq:bayes}). This completes the derivation.

We now illustrate the rule for a system of $M+N$ qubits, for which the
Hilbert space ${\cal H}$ of each subsystem is two-dimensional. An
arbitrary exchangeable state of $M+N$ qubits can be written in the
form
\begin{equation}
\R{M+N} = \int\!\!\int\!\!\int dx\,dy\,dz\,p(x,y,z)\rxyz^{\otimes(M+N)} \;,
\end{equation}
where $\rxyz={1\over2}(\hat 1+x\sx+y\sy+z\sz)$ and the integrals range
over the volume of the sphere of radius 1. Here $\sx$, $\sy$, $\sz$
are the Pauli operators, and $\hat 1$ denotes the unit operator.

Now assume that $\sz$ measurements are performed on $M$ qubits. The
probability of obtaining the result $\pm1$, given state $\rxyz$, in a $\sz$
measurement on a single qubit is
\begin{equation}
p(\pm1|\rxyz) = {1\over2}(1\pm z) \;.
\end{equation}
If the $M$ measurements of $\sz$ yield $M_+$ results of +1 and $M_-$ results 
of $-1$, where $M_++M_-=M$, then the state of the remaining $N$ qubits is
\begin{equation}
\R N_{M_+,M_-}=
\int\!\!\int\!\!\int dx\,dy\,dz\,p(x,y,z|M_+,M_-)\rxyz^{\otimes N}\;,
\label{eq:spinpost}
\end{equation}
where 
\begin{eqnarray}
p(x,y,z|M_+,M_-) &=& {\cal N} p(x,y,z) \nonumber\\
&& \times \left({1+z\over2}\right)^{M_+} 
\left({1-z\over2}\right)^{M_-} \;,
\end{eqnarray}
${\cal N}$ being a normalization factor. 

In the limit $M\rightarrow\infty$, assuming $(M_+-M_-)/M\rightarrow E_z$, 
we obtain
\begin{equation}
p(x,y,z|M_+,M_-) \rightarrow 
p(x,y|E_z) \delta(z-E_z)\;,
\label{eq:limit}
\end{equation}
where $p(x,y|E_z)=p(x,y,E_z)/\int\!\!\int dx\,dy\,p(x,y,E_z)$ is the prior 
conditional probability for $x$ and $y$, given that $z=E_z$.  
Equation~(\ref{eq:limit}) expresses clearly the gain in information
about $z$.  For an isotropic prior,
\begin{equation}
  \label{eq:isotropic}
  p(x,y,z)=p\Bigl(\sqrt{x^2+y^2+z^2}\Bigr) \;,
\end{equation}
the marginal state for a single subsystem before any measurements is
the maximally mixed state $\R1={1\over2}\hat 1$.  After $M$ measurements 
of $\sz$, in the limit $M\rightarrow\infty$, the marginal state for a 
single additional subsystem is  
\begin{equation}
\R1_{E_z} = {1\over2} (\hat 1 + E_z\sz) \;,
\label{eq_one_spin}
\end{equation}
which is the state obtained in \cite{Buzek1998b}. Our analysis puts this 
in a clear perspective: the data dictate the expectation 
value $\langle\sz\rangle=E_z$ for the state~(\ref{eq_one_spin}); 
for an isotropic prior, the $\sz$ measurements tell one nothing about the
direction of the spin in the $x$-$y$ plane, so $\sx$ and $\sy$ retain
the zero expectation values that apply to the prior marginal state of 
a single subsystem.

It is important to note that the state $\R1_{E_z}$ does not allow one to make
predictions about frequencies in future repeated measurements of, e.g., the
observable $\sx$. Although $\tr\big(\R1_{E_z}\sx\big)=0$, it would be wrong 
to predict that the frequency of the outcome +1 in a large number of future 
$\sx$ measurements will be close to 1/2. The correct prediction for 
future $\sx$ measurements follows from the full state $\R N$ with the limiting 
posterior (\ref{eq:limit}); for the probability of obtaining $N_+$ results 
of $+1$ and $N_-$ results of $-1$ in $N$ measurements of $\sx$, we get
\begin{eqnarray}
p(N_+,N_-|E_z) = 
{N!\over N_+!N_-!} \int\!\!\int dx\,dy\, p(x,y|E_z) 
\left({1+x\over2}\right)^{N_+} \left({1-x\over2}\right)^{N_-} .
\label{x_spin_prob}
\end{eqnarray}
Only in the extreme case that the prior has the special form
$p(x,y,z)=p(y,z)\delta(x)$ does the probability~(\ref{x_spin_prob}) become
identical to the prediction $P(N_+,N_-)=2^{-N}N!/N_+!N_-!$ that would follow
from assigning the product state $\r^{(1)\otimes N}_{E_z}$ to the $N$ subsystems.
It is clear that this prediction is not implied by the $\sz$ measurement data
and is therefore unwarranted unless there is additional prior information.

The marginal state $\R1_{E_z}$ of Eq.~(\ref{eq_one_spin}) can also be derived 
from the principle of maximum entropy (MAXENT)
\cite{Jaynes1957a,Jaynes1957b}.  If all that is known about the state $\r$ of
some system is the expectation value of one or several observables, the MAXENT
state assignment results from maximizing the von Neumann entropy of $\r$
subject to the constraints given by the expectation values (see
Ref.~\cite{Balian1987} for a derivation of the MAXENT principle in the
quantum case).

In the example above, the MAXENT assignment following from the constraint
$\langle\sz\rangle=E_z$ for a single subsystem is identical to the marginal 
state (\ref{eq_one_spin}). This identity has also been noted by Bu\v{z}ek 
{\it et al.} \cite{Buzek1998b}, who state that ``\,\ldots\ as soon as the 
number of measurements becomes large then [the] Bayesian inference scheme 
becomes equal to the reconstruction scheme based on the Jaynes principle 
of maximum entropy \ldots\,.'' This statement is misleading, however, 
since the equality holds only for the marginal state of a single subsystem 
(and even then only under the isotropy assumption~(\ref{eq:isotropic}) 
for the prior). Unlike the full state $\R N_{M_+,M_-}$ in 
Eq.~(\ref{eq:spinpost}), found via Bayes's rule, the single-subsystem 
state $\R 1_{E_z}$ derived via MAXENT does not allow one to make predictions 
for measurements on more than one subsystem.

On the other hand, applying MAXENT directly to $N$ subsystems fails for 
the following reason, well known from classical probability theory
\cite{Jaynes1986,Skilling1989}. Maximizing the von Neumann entropy of 
$\R N$ subject to the constraint that $\langle\sz\rangle=E_z$ for each
subsystem yields the product state 
$\R N_{\rm MAXENT}=\r^{(1) \otimes N}_{E_z}$.  As discussed above, this 
state assignment is unwarranted because it leads to predictions for, say, 
future $\sx$ measurements which are in no way implied by the constraint 
on $\langle\sz\rangle$.  Furthermore, any product state assignment precludes 
learning from subsequent measurements, even though that should be possible, 
as was discussed in the paragraph after Eq.~(\ref{eq:productprior}).

If the measurements on individual subsystems correspond to an informationally
complete POVM \cite{Kraus1983} or if they contain sequences of measurements
of a tomographically complete set of observables \cite{Buzek1998b}, the 
posterior probability on density operators approaches a $\delta$ function in 
the limit of many measurements.  This is the case of quantum state tomography 
\cite{Vogel1989b,Smithey1993,Leonhardt1995}, which can thus be viewed as 
a special case of quantum Bayesian inference. In this limit, the exact form 
of the prior probability on density operators becomes irrelevant. In all 
other situations, however, there will be some dependence on this prior.

TAB thanks Bob Griffiths and Oliver Cohen for helpful discussions.  TAB 
was supported in part by NSF Grant No. PHY-9900755, and CMC received partial 
support from ONR Grant No.~N00014-93-1-0016.

%\bibliographystyle{prsty}
%\bibliography{/home/rschack/lit/p}

\begin{thebibliography}{10}

\bibitem{Bernardo1994}
J.~M. Bernardo and A.~F.~M. Smith, {\em Bayesian Theory} (Wiley, Chichester,
  1994).

\bibitem{Malakoff1999}
D. Malakoff, Science {\bf 286},  1460  (1999).

\bibitem{Sykora1974}
S. S\'ykora, J. Stat.\ Phys.\ {\bf 11},  17  (1974).

\bibitem{Helstrom1976}
C.~W. Helstrom, {\em Quantum Detection and Estimation Theory} (Academic Press,
  New York, 1976).

\bibitem{Holevo1982}
A.~S. Holevo, {\em Probabilistic and Statistical Aspects of Quantum Theory}
  (North Holland, Amsterdam, 1982).

\bibitem{Jones1991a}
K.~R.~W. Jones, Ann.\ Phys.\ {\bf 207},  140  (1991).

\bibitem{Vogel1989b}
K. Vogel and H. Risken, Phys.\ Rev.\ A {\bf 40},  2847  (1989).

\bibitem{Smithey1993}
D.~T. Smithey, M. Beck, M.~G. Raymer, and A. Faridani, 
Phys.\ Rev.\ Lett.\ {\bf 70},  1244  (1993).

\bibitem{Leonhardt1995}
U. Leonhardt, Phys.\ Rev.\ Lett.\ {\bf 74},  4101  (1995).

\bibitem{Massar1995}
S. Massar and S. Popescu, Phys.\ Rev.\ Lett.\ {\bf 74},  1259  (1995).

\bibitem{Hradil1997}
Z. Hradil, Phys.\ Rev.\ A {\bf 55},  R1561  (1997).

\bibitem{Slater1997}
P.~B. Slater, J. Math.\ Phys.\ {\bf 38},  2274  (1997).

\bibitem{Derka1998}
R. Derka, V. Bu\v{z}ek, and A.~K. Ekert, Phys.\ Rev.\ Lett.\ {\bf 80},  1571
  (1998).

\bibitem{Buzek1998b}
V. Bu\v{z}ek, R. Derka, G. Adam, and P.~L. Knight, 
   Ann.\ Phys.\ {\bf 266},  454 (1998).

\bibitem{Tarrach1999}
R. Tarrach and G. Vidal, Phys.\ Rev.\ A {\bf 60},  3339  (1999).

\bibitem{Banaszek2000}
K. Banaszek, G.~M. D'Ariano, M.~G.~A. Paris, and M.~F. Sacchi, Phys.\ Rev.\ A
  {\bf 61},  10304  (2000).

\bibitem{Gill2000a}
R.~D. Gill and S. Massar, Phys.\ Rev.\ A {\bf 61},  42312  (2000).

\bibitem{Lemm2000a}
J.~C. Lemm, J. Uhlig, and A. Weiguny, Phys.\ Rev.\ Lett.\ {\bf 84},  2068
  (2000).

\bibitem{Paini-0002}
M. Paini, quant-ph/0002078.

\bibitem{Kraus1983}
K. Kraus, {\em States, Effects, and Operations. Fundamental Notions of Quantum
  Theory} (Springer, Berlin, 1983), Lecture Notes in Physics Vol.\ 190.

\bibitem{Stormer1969}
E. St{\o}rmer, Journal of Functional Analysis {\bf 3},  48  (1969).

\bibitem{Hudson1976}
R.~L. Hudson and G.~R. Moody, Z. Wahrscheinlichkeitstheorie verw.\ Geb.\ {\bf
  33},  343  (1976).

\bibitem{DeFinetti1990}
B. de~Finetti, {\em Theory of Probability} (Wiley, New York, 1990).

\bibitem{Caves2000a}
C.~M. Caves, C.~A. Fuchs, and R. Schack, (2000), in preparation.

\bibitem{Bures1969}
D. Bures, Trans.\ Am.\ Math.\ Soc.\ {\bf 135},  199  (1969).

\bibitem{Braunstein1994a}
S.~L. Braunstein and C.~M. Caves, Phys.\ Rev.\ Lett.\ {\bf 72},  3439  (1994).

\bibitem{Zyczkowski1998b}
K. \.Zyczkowski, P. Horodecki, A. Sanpera, and M. Lewenstein, Phys.\ Rev.\ A
  {\bf 58},  883  (1998).

\bibitem{Jaynes1957a}
E.~T. Jaynes, Phys.\ Rev.\ {\bf 106},  620  (1957).

\bibitem{Jaynes1957b}
E.~T. Jaynes, Phys.\ Rev.\ {\bf 108},  171  (1957).

\bibitem{Balian1987}
R. Balian and N.~L. Balazs, Ann.\ Phys.\ {\bf 179},  97  (1987).

\bibitem{Jaynes1986}
E.~T. Jaynes,  in {\em Maximum Entropy and Bayesian Methods in Applied
  Statistics}, edited by J.~H. Justice (Cambridge University Press, Cambridge,
  1986), p.\ 26.

\bibitem{Skilling1989}
J. Skilling,  in {\em Maximum Entropy and Bayesian Methods}, edited by J.
  Skilling (Kluwer Academic Publishers, Dordrecht, The Netherlands, 1989), 
  p.\ 45.

\end{thebibliography}

\end{document}